\documentclass[manuscript]{emulateapj}
\usepackage{natbib}
\usepackage{graphicx}
\usepackage{multirow}
\usepackage{txfonts}

\shorttitle{Accretion dynamics of H 1743-322 from TCAF Solution}
\shortauthors{S. Mondal, D. Debnath, and S. K. Chakrabarti}

\begin{document}

\title{Inference on accretion flow dynamics using TCAF solution from the analysis of spectral evolution of H 1743-322 during 2010 outburst}

\author{Santanu Mondal\altaffilmark{1}, Dipak Debnath\altaffilmark{1}, Sandip K. Chakrabarti\altaffilmark{2,1}}
\altaffiltext{1}{Indian Center for Space Physics, 43 Chalantika, Garia St. Rd., Kolkata, 700084, India.}
\altaffiltext{2}{S. N. Bose National Centre for Basic Sciences, Salt Lake, Kolkata, 700098, India.}

\email{santanu@csp.res.in; dipak@csp.res.in; chakraba@bose.res.in}

\date{Received: 2014 January 16; Accepted: 2014 March 5}

\begin{abstract}

We study accretion flow dynamics of Galactic transient black hole candidate (BHC) H~1743-322 
during its 2010 outburst by analyzing spectral data using Two Component (Keplerian and sub-Keplerian) 
Advective Flow (TCAF) solution, after its inclusion in XSPEC as a local model. We compare our TCAF solution 
fitted results with combined disk black body and power-law model fitted results and find a similar smooth 
variation of thermal (Keplerian or disk black body) and non-thermal (power-law or sub-Keplerian) fluxes/rates 
in two types of model fits. For a spectral analysis, $2.5-25$~keV spectral data from RXTE PCA instrument 
are used. From the TCAF solution fit, accretion flow parameters, such as Keplerian rate, 
sub-Keplerian rate, location of centrifugal pressure supported shock and strength of the shock 
are extracted, thus providing a deeper understanding of accretion process and properties of accretion 
disks around BHC H~1743-322 during its X-ray outburst. Based on the halo to disk accretion rate ratio (ARR), 
shock properties, accretion rates and nature of quasi-periodic oscillations (QPOs, if observed)
entire outburst is classified into four different spectral states, such as, {\it hard, hard-intermediate, 
soft-intermediate}, and {\it soft}. From time variation of intrinsic flow parameters
it appears that their evolutions in decline phase do not retrace path of rising phase. Since our current model
does not include magnetic fields, spectral turnover at  energies beyond 500-600 keV cannot be explained.
\end{abstract}

%\begin{keywords}
%{Stars:individual (H~1743-322), X-ray sources, Black Holes, accretion disks, shock waves, Spectrum, Radiation hydrodynamics}
%\end{keywords}

\section{Introduction}

Galactic transient black hole candidates (BHCs) are very interesting objects to study because these 
sources generally show evolutions in their temporal and spectral properties during their outburst phases, 
which are strongly correlated to each other. In last two decades, especially after the launch of 
{\it Rossi X-ray Timing Explorer} (RXTE), our understanding on BHCs 
has progressed significantly, but not to the extent that we can visualize how the 
flow configuration and properties are changing within short time scales, say, 
less than a day.  In general, it has been found that during outbursts of a BHC, four basic spectral states, 
namely, {\it hard, hard-intermediate, soft-intermediate}, and {\it soft} are observed 
(see,  Nandi et al., 2012 and references therein). % \citep[see,][and references therein]{Nandi12}.
% McClintok \& Remillard, 2003; Nandi et al., 2012). Detailed discussions about the spectral states 
%transition is in literature (Homan \& Belloni, 2005a; Nandi et al., 2012). 
Detailed discussions on the evolution of the temporal and spectral properties of different BHCs during 
their outbursts were made by several groups %\citep[see,  e.g.,][]{MR06,Belloni05,Nandi12}. 
(see for e.g., McClintok \& Remillard, 2006; Belloni et al., 2005, Nandi et al., 2012). 
Different branches of hardness intensity diagram (HID) %or the so-called `q' diagram 
(Maccarone \& Coppi, 2003; Belloni et al., 2005, etc.) %\citep{Maccarone03,Belloni05} 
are also found to be related to different spectral states. 
%In this {\it paper}, we study the important physical properties, such as the time variation of the accretion 
%rates of the Keplerian and sub-Keplerian components, the shock (when present) locations and strengths of 
%H~1743-322 during its 2010 outburst by analyzing the spectral properties using Two Component Advective Flow 
%(TCAF) solution (Chakrabarti \& Titarchuk, 1995, hereafter CT95).%\citep[][hereafter CT95]{CT95}.

It is well established that a standard Keplerian disk (Shakura \& Sunyaev, 1973)%\citep{SS73} 
cannot explain full X-ray spectrum from black hole candidates and one necessarily requires a second 
component, namely, the so-called `Compton' cloud (Sunyaev \& Titarchuk, 1980, 1985), %\citep{ST80,ST85}, 
to produce the power-law part of the spectrum. There are speculations regarding origin and nature of 
this Compton cloud which range from a magnetic corona (Galeev, Rosner \& Viana, 1979) %\citep{Galeev79} 
to a hot gas corona over the disk (Haardt \& Maraschi, 1993; Zdziarski et al., 2003). %\citep{Haardt93,Zdziarski03}.
Observational evidences show that while both the components must be dynamic, one component moves faster (e.g., Smith, 
et al. 2002; Soria, et al., 2001; Wu, et al. 2002; Cambier \& Smith, 2013) %\citep[e.g.,][]{Soria01,Smith02,Wu02,Cambier13} 
very much like the low-angular momentum, sub-Keplerian transonic flow component as incorporated 
in Chakrabarti \& Titarchuk (1995, hereafter CT95). In Two Component Advective Flow (TCAF) 
solution, Chakrabarti and his collaborators, 
even before the RXTE was launched, envisaged that Compton cloud is actually the inefficiently radiating 
transonic flow (Chakrabarti, 1990) %citep{C90}
having very low, or {\it sub-Keplerian} angular momentum. Matter in this accretion flow becomes hot close to 
the black hole where the centrifugal pressure starts dominating and an accretion shock may or may not form 
depending on whether or not the Rankine-Hugoniot shock conditions are satisfied 
(CT95; Chakrabarti, 1997, hereafter C97). %\citep[CT95;][hereafter C97]{C97}.
Recently, Mondal \& Chakrabarti (2013), and Giri \& Chakrabarti (2013) %\citet{SM13} and \citet{GC13} 
showed that a self-consistent and stable transonic solution exists which supports the solution envisaged by CT95.

The stellar mass BHC H~1743-322 is very intriguing, because in the last decade after its 
re-discovery in 2003 (Revnivtsev et al. 2003), %\citep{Revnivtsev03},
it showed several X-ray outbursts in regular intervals of $1-2$ years. This source was in a quiescent state 
for a long time - only a couple of X-ray activities by EXOSAT in 1984 (Reynolds, 1999) %\citep{Reynolds99} 
and by TTM/COMIS onboard Mir-Kvant in 1996 (Emelyanov et al., 2000) %\citep{Emelyanov00} 
were reported after its first detection in Aug-Sep, 1977, 
with Ariel-V All-Sky Monitor (Kaluzienski \& Holt, 1977) %\citep{Kaluzienski77}
and HEAO-1 satellite (Doxsey et al., 1997). %\citep{Doxsey77}.
This Low Mass X-ray Binary (LMXB) system is located at R.A. = $17^h46^m15^s.61$ 
and Dec.=$-32^\circ14'00''.6$ (Gursky et al., 1978). %\citep{Gursky78}. 
Mass of this BHC has not been dynamically confirmed yet, although (P\'{e}tri, 2008) %\citet{Petri08} 
predicted that its mass is in between $9 M_\odot$ to $13 M_\odot$ with their 
%{(\bf McClintok paper M = 11, our paper M = $11.4 \pm ???$ )} \citep{DD14}.
high frequency QPO model. Steiner et al. (2012) % \citet{Steiner12}
has confirmed its distance $D$ = $8.5\pm0.8$~kpc, disk inclination angle $\theta = 75^\circ\pm3^\circ$,
and spin ($-0.3<a_*<0.7$ with a 90\% confidence).

%During its 2003 outburst, the low-frequency as well as high frequency quasi-periodic oscillations (QPOs) 
%are observed with RXTE/PCA (Homan et al. 2005; Remillard et al. 2006) %\citep{Homan05, Remillard06}
%data. Also in 2003 outburst, strong spectral variability (Capitanio et al. 2005) %\citep{Capitanio05}
%was observed. In 2003, another important discovery was the large-scale relativistic 
%radio jet (Rupen et al. 2004; Corbel et al. 2005). 

Recently, in 2010, H~1743-322 was again found to be active in X-rays (Yamaoka et al., 2010) %\citep{Yamaoka10} 
with a similar characteristics of temporal and spectral evolutions as observed in other transient BHCs 
(Debnath et al., 2008, 2010; Nandi et al., 2012 and references therein). %\citep[][and references therein]{DD08,DD10,Nandi12}. 
The outburst was observed for a short time period of around two months and 
RXTE had a full coverage of this source on a daily basis. 
In Debnath et al. (2013; hereafter Paper-I), %\citet[][hereafter Paper-I]{DD13}, 
a detailed study of the temporal and spectral properties of the source during its two successive outbursts 
(2010 \& 2011) using RXTE/PCA archival data are presented. In Paper-I, the spectral properties of the source 
were studied with a combination of conventional thermal (disk black body) and non-thermal (power-law) model 
components. In order to understand more realistic picture of the accretion flow dynamics, one needs to study 
spectral properties with a more physical model (such as TCAF), which would enable one to extract actual 
physical parameters of the accretion flow. 
In this {\it paper}, we study important physical properties, such as time variation of accretion 
rates of Keplerian and sub-Keplerian components, shock (when present) location and strength  of the flow around
H~1743-322 during its 2010 outburst by analyzing spectral properties using TCAF solution. 
Based on spectral classification method as defined in Debnath, Mondal \& Chakrabarti (2014a), %\citet{DD14a}, 
four basic spectral states %(hard, hard-intermediate, soft-intermediate, soft) 
are also observed during the entire period of this outburst; which also form a hysteresis loop, similar to the HID, 
where different spectral states belong to different branches of diagram (see. Fig. 4).

This {\it Paper} is organized in the following way: in the next Section, we discuss observation 
and data analysis procedures using HEASARC's HeaSoft software package. In \S 3, we present results 
of spectral analysis using TCAF based model {\it fits} file and variation of different flow parameters 
with observational results. Finally, in \S 4, we present a brief discussion and make our concluding remarks.

\section{Observation and Data Analysis}

We analyze data of $26$ observational IDs starting from 2010 August 9 (Modified Julian Day, i.e., MJD = 55417) 
to 2010 September 30 (MJD = 55469). 
We carry out data analysis using FTOOLS software package HeaSoft version HEADAS 6.12 and XSPEC version 12.7. For generation 
of source and background `.pha' files and spectral fitting using TCAF solution, we use same method as described in 
Debnath, Mondal \& Chakrabarti (2014a, hereafter DMC14), and Debnath, Chakrabarti \& Mondal (2014b, hereafter DCM14). 
%(\citealt{DD14a}, hereafter DMC14; \citealt{DD14b}, hereafter DCM14).

The $2.5-25$ keV PCA background subtracted spectra are fitted with TCAF based model {\it fits} file. To achieve 
the best fit, a Gaussian line of peak energy around $6.5$~keV (iron-line emission) is used. For the entire outburst, 
we keep hydrogen column density (N$_{H}$) fixed at 1.6$\times$~10$^{22}$~atoms~cm$^{-2}$ for absorption 
model {\it wabs} and assume a $1.0$\% systematic error (Paper-I). 
%In the entire PCA data analysis, we include the dead-time correction and also the pca breakdown correction 
%(because of the leakage of propane layers of PCUs). 
After achieving best fit based on reduced chi-square value ($\chi^2_{red} \sim 1$), `err' command 
is used to find 90\% confidence error values for the model fitted parameters. 
We have not included HEXTE data since the rocking mechanism stopped in 2010 and thus it is difficult to 
subtract the background.

For a spectral fit using the TCAF based model, one needs to supply a total of six input parameters: 
$i)$ Keplerian rate ($\dot{m_d}$ in Eddington rate), $ii)$ sub-Keplerian rate ($\dot{m_h}$ in Eddington rate),
$iii)$ black hole mass ($M_{BH}$) in solar mass ($M_\odot$) unit, $iv)$ location of the shock ($X_s$ in 
Schwarzschild radius $r_g$=$2GM/c^2$), $v)$ compression ratio ($R$) of the shock, and $vi)$ the model 
normalization value ($norm$) of %%$\frac{1}{4\pi D^2} cos(i)$, 
$\frac{R_z}{4\pi D^2} sin(i)$, where `$R_z$' is the effective height of the Keplerian component in $Km$ 
at the pre-shock region,`$D$' is the source distance in $10$~kpc unit and `$i$' is the disk inclination angle. 
In order to fit a black hole spectrum with TCAF model in XSPEC, we generate model {\it fits} file ({\it TCAF.fits}) 
using theoretical spectra generating software by varying five input parameters in CT95 code and then include it 
in XSPEC as a local additive model. Detail description of the range of input parameters and generation procedure 
is mentioned in DCM14 and DMC14. In this Paper, for the spectra of entire outburst data, we consider mass of the 
black hole as 11.4$\pm$1.9$M_\odot$ (Debnath et al., 2014c). %\citep{DD14c}. 

%Although the mass of the black hole has not yet been measured dynamically and can fall in between 9$M_\odot$ 
%to 13$M_\odot$ (P\'etri 2008). Here we consider the mass as 11.4$M_\odot$, as we have seen from our 
%calculation, which will be published elsewhere \citep{DD14c}. %(Debnath et al., 2014c).

\section{Results}

Accretion flow dynamics during an outburst phase of transient BHCs can be well understood
by model analysis of spectral and temporal behavior of the source. These behaviors have already 
been discussed in Paper-I though the analysis was made using a combination of disk black body (DBB) and 
power-law (PL) model components which give only the gross properties of the disk. TCAF based model 
goes one step further in extracting the detailed flow parameters, such as two disk rates and shock properties.
Furthermore, this makes the boundary of the states more well-defined.
Thus, for accretion dynamics, we need to use the TCAF based model {\it fits} file.

\subsection{Results of Spectral Data Fitted by TCAF Solution}

Figure 1 shows variation of X-ray intensities, QPO frequencies along with model fitted parameters. In Fig. 1a, 
variation of background subtracted RXTE PCA count rate in $2-25$~keV ($0-58$ channels) energy 
band with day (MJD) is shown. Figs. 1b \& 1c show variations of combined DBB and PL model fitted total 
spectral flux (flux contributions for DBB and PL model components are calculated by using convolution 
model `cflux' technique after fitting spectra with combined model components) in $2.5-25$~keV energy band 
and TCAF model fitted total accretion rates (combined Keplerian disk and sub-Keplerian halo rates) in the same 
energy band. In Fig. 1d, variation of {\it Accretion Rate Ratio} (ARR, defined to be 
ratio of sub-Keplerian halo rate $\dot{m_h}$ and Keplerian disk rate $\dot{m_d}$).
Variations of TCAF model shock locations ($X_s$) and compression ratios are shown in Fig. 1e \& 1f respectively. 
Observed QPO frequencies are shown in Fig. 1g. Depending on variation of ARR and nature 
(shape, frequency, $Q$ value, rms\% etc) of QPOs (when observed), four different spectral states 
such as {\it hard (HS), hard-intermediate (HIMS), soft-intermediate (SIMS), soft (SS)} could be identified 
during the entire outburst of H~ 1743-322 (for details see, DMC14). We observe that spectral transitions 
between these states occur approximately at the same day as reported in Paper-I (marked with vertical 
dashed lines in Figs. 1-2), where spectral classifications were done on the basis of degree of importance of 
DBB and PL model components (fluxes) and properties of QPOs. 
We see that the TCAF model fits well for all four spectral states (model fitted $\chi^2_{red}$ varies $\sim 0.9-2$) 
obeying spectral results of combined DBB \& PL. As of DMC14 for 2010-11 GX~339-4 outburst, here we also find 
two surges in ARR at the boundaries of hard-intermediate states, i.e., from hard to hard-intermediate state 
during rising phase or from hard-intermediate to hard state during declining phase of the outburst) transition days. 
Note that from the TCAF fit, sum of the accretion rates appears to be almost constant in the hard states, 
both before and after the outburst. At the onset of the soft-intermediate state, the total rate rises very sharply.
This continues till soft-intermediate state ends. The rate decays monotonically till the end of 
the hard intermediate state. This will be clear from Fig. 2, where we show comparative variations 
of the combined DBB and PL model fitted DBB and PL fluxes along with TCAF model fitted Keplerian disk 
($\dot{m_d}$) and sub-Keplerian halo ($\dot{m_h}$) rates with day (MJD). 
It is important to note that these two variations are going hand in hand. 
We shall return back to discussions on Figs. 1(e-g) later. %%%%%%%%%%%%%%%%%%%%%%

As a further consistency check of TCAF, we may ask the following:  
since the power-law component is due to scattering of intercepted black body photons
by Compton cloud, which, according to TCAF, is produced by radiatively inefficient, 
low angular momentum transonic halo component, could variations of PL flux and 
$\dot{m_h}$ be similar? From the Figure 2(c-d) the similarity is obvious. Indeed the behavior shown in 
Fig. 2 justifies us using the more physical model, i.e., TCAF 
which directly gives the intrinsic flow parameters. Furthermore, this 
also shows that the halo rate plays a major role in deciding the spectral properties of the flow. 
However, at times, the two rates vary independently as is obvious from Fig 1d. This will be discussed later.

%In Figs. 4(a-f), TCAF model fitted $2.5-25$~keV background subtracted PCA spectra with variation of 
%$\Delta \chi$ in four different (hard, hard-intermediate, soft-intermediate, soft) spectral states (for 
%both rising and declining phases), whose parameters are presented in Table 1.
In Table 1, TCAF model fitted parameters along with frequency of primary observed QPOs (if present)
are mentioned. In Fig. 3, unabsorbed theoretical model spectra for seven states in $0.005-1000$~keV energy 
range, selected from different spectral states of the outburst as marked in the Table, are shown. Note that 
relatively harder states clearly show spectral turnover above $\sim 300$keV. It is also to be noted that since
present model of TCAF does not include magnetic fields explicitly, inverse Comptonization of non-thermal 
photons produced in the post-shock region could not be included and thus turnovers which may occur at much 
higher energies in sources such as Cyg X-1 (Zdziarski 2000; Zdziarski et al. 2001; Chakrabarti \& Mondal, 2006) 
cannot be fitted with the fits file generated using the current TCAF model.
 
\subsection{Evolution of Spectral and Temporal Properties during the Outburst}

Detailed temporal and spectral properties of this candidate are discussed by several authors on the basis 
of whether QPO observation was made or not %\citep[][Paper-I]{Belloni05,McClintock09}. 
(Belloni et al. 2005; McClintock et al. 2009; Paper-I).  
However, since we have physical parameters on each day, it may be instructive to 
check if there is any physical way to differentiate one spectral state from another.
In Figs. 1d and 1g, we see variations of ARR and QPO frequencies with time 
in MJD. It seems that ARR, total flow/accretion rate ($\dot{m_d}$+$\dot{m_h}$), shock locations,
compression ratios, etc. in conjunction with QPOs provide a better understanding 
on the classification of spectral states.

\noindent{\it (i) Hard State in the Rising phase:}
For the first $3$ days of RXTE observation (from MJD = 55417.29 to 55419.11), source was in a hard state with 
increasing total flux and non-thermal PL flux (or, equivalently, 
sub-Keplerian halo rate; see, Figs. 1-2). 
During this phase, halo rate is increasing faster than disk rate as the infall time of the
halo is shorter than that of the disk. ARR is monotonically increasing and reached a value of $3.33$, maximum
for the entire outburst. QPO frequencies are observed to be increasing monotonically 
from $0.919$~Hz to $1.045$~Hz (see, Paper-I). 
We define the day of maximum ARR as the transition day from the hard to hard-intermediate spectral state. 
%On 2010 March 22 (MJD = 55277.48), a {\it kink} is observed in ARR plot (Fig. 1d),
%i.e. a sudden rise in halo rate compared to the disk rate is occurred, where the first prominent QPO of frequency
%$0.102$~Hz is observed. 
%After that QPO frequency is increased monotonically. 

\noindent{\it (ii) Hard-Intermediate State in the Rising phase :}
The source was in this state for the next $\sim 5$ days after the transition day (MJD = 55419.11). During this
phase, QPO frequency continues to increase monotonically till it reached from $1.045$~Hz on MJD = 55419.11
to $4.796$~Hz on MJD = 55424.06. Supply of sub-Keplerian matter is continued at the same rate
and a part of it is converted to Keplerian matter due to viscous effects, thereby increasing its rate.
Consequently, ARR is decreased. We define the end of hard-intermediate state 
up to time when the episode of constant ARR begins.

\noindent{\it (iii) Soft-Intermediate State in the Rising phase :}
The constancy of ARR lasted till total rate suddenly jumps. Both disk and halo rates along with 
PCA count rapidly increased till QPOs last. On this day (MJD = 55425.16), the observed QPO frequency 
($3.558$~Hz) did not increase as before. This date of the highest total rate and total PCA counts ends the 
soft-intermediate state and usher the object to the next, i.e., soft state.

\noindent{\it (iv) Soft State:}
The source is observed at this state for the next $\sim 26$ days (up to MJD = 55450.34), where the spectra 
were mostly dominated by thermal photons. Shock location is very close to the black hole, but is not
oscillating as the resonance oscillation condition is not fulfilled in presence of rapid cooling. As a result, the
QPO is absent. The continuous drainage of Keplerian and sub-Keplerian matter reduced the total rate monotonically,
while keeping ARR roughly constant. This state continues till the drainage is just enough to bring back 
resonance condition for the oscillation of the CENtrifugal pressure supported Boundary Layer (CENBOL) 
whose outer boundary is the shock (see, Molteni, Sponholz and Chakrabarti, 1996; hereafter MSC96) %citep{MSC96}
and QPOs reappear. ARR became almost same as that on the date
QPO was last observed in the Soft intermediate state. On MJD = 55451.17, a state transition is observed. 

\noindent{\it (v) Soft-Intermediate State in the Declining phase  :}
For the following $\sim 5$ days, the source was observed in this state. During this period, the total flux 
(DBB+PL, PCA count rate and $\dot{m_d}$+$\dot{m_h}$) is almost constant, though individually the 
rates fluctuate. Sporadic QPOs of $\sim 2$ Hz are observed during this spectral state.
Sporadic phase of QPO is over on MJD = 55455.44, which signifies the end of this state.

\noindent{\it (vi) Hard-Intermediate State in the Declining phase  :}
On MJD = 55455.44, hard-intermediate state starts. After that, continuous QPOs are observed. ARR
on this day is roughly the same it was in the last day of hard-intermediate state in the rising phase.  
Initially, the Keplerian disk is drained more rapidly and thus ARR increases rapidly.
Total accretion rate, net PCA count etc.  continued to be drained and 
centrifugal pressure supported shock location started receding (Fig. 1e) as the 
incoming flow pressure drops. As a result, QPO frequency drops steadily 
(see, Paper-I). On MJD = 55462.56, maximum ARR (=2.62) of 
declining phase is observed, which defines the transition day from hard-intermediate state to hard state. 

\noindent{\it (vii) Hard State in the Declining phase:}
Source is observed in this spectral state till the end of the observation of 2010 outburst. In this state 
ARR value decrease monotonically from $2.62$ to $1.50$ and observed QPOs also decrease monotonically 
from $0.74$ Hz to $79$ mHz (see, Paper-I). However, the total rate becomes roughly constant as in the 
rising phase. The rising phase was observed several days after it began. In the declining phase observation 
continued till PCA count was much lower than that on the first day of the rising phase.
The end of hard state concludes the outburst.

\section{Discussions and Concluding Remarks}

We analyzed spectral properties of Galactic transient black hole candidate H~1743-322 during its 2010 outburst 
using two component advective flow (TCAF) solution based model after its inclusion as a local additive table odel 
in HEASARC's spectral analysis software package XSPEC. We generate $\sim 4\times10^5$ model spectra using 
CT95 code to fit with the observational data. Our fits file generation procedure is the same as that presented 
in DMC14. Flow parameters (Keplerian and sub-Keplerian accretion rates) extracted from TCAF model 
spectral fit generally match with DBB and PL fluxes (see, Figs. 1-2). Total rate, shock location, QPOs 
in different spectral states have been observed to behave very reasonably. From these quantities, we have a 
complete physical picture of when exactly spectral state transitions occur. 

In Fig. 4(a-b) we give two Figures where we show variation of derived physical quantities
of the flow from our fits. In (a), we plot variation of PCA count rate as a function of ratio 1/ARR and  in (b),
disk rate as a function of the halo rate for the whole outburst. B, C, D, E, F and G are points where state 
transitions take place. In (a), segments A-B (black online) and G-H (orange online) show highest ARR locations 
though PCA counts are different, the rising phase A-B (when first caught) being considerably brighter. 
Segments B-C (red online) and G-F (cyan online) have roughly similar range of ARR. The pattern of 
first horizontal and then vertical variation of PCA count in the rising phase is reversed in the decline phase. 
Segments C-D (green online) and E-F (magenta online) are soft-intermediate states occurring roughly at the 
same ARR value in the rising phase (sharp rise in PCA count) but rapid oscillation in ARR, though ARR at the 
edges remain roughly the same. Segment D-E (blue online) is fluctuating, drainage of disk and 
halo being alternately high. Fig. 4b clearly shows that the two rates are independent and they 
do not always increase or decrease together, except only in soft state. 
Indeed, sometimes they are horizontal and sometimes vertical. 
In these cases, only one component changes keeping the other component as constant.

As far as the low frequency QPO behavior is concerned, it can occur only when the cooling time scale roughly agrees
with the infall time scale (MSC96). In hard and hard-intermediate states this is easily achievable as shocks are formed 
and Compton cooling time scale is similar to infall time scale from the shock to the inner edge of the disk. 
In soft and soft-intermediate states (when ${\dot m}_d > 1$, see Fig. 4b), CENBOL is cooled down very rapidly 
and shock may or may not be seen. This is true in both rising and decline phases. In soft-intermediate states 
QPOs are therefore seen sporadically. When the total rate is maximum, soft intermediate state transits 
to soft state and QPO is not seen any more. In hard intermediate states, ${\dot m}_d$ decreases keeping while 
${\dot m}_h$ remains roughly a similar number. Thus ARR increases while going from hard-intermediate states to hard 
states. ARR develops a sharp maximum on the day of transition from hard to hard-intermediate state in the rising 
phase, and the reverse in the decline phase. This does show that QPOs are strongly coupled to cooling properties 
of the disk. 

In the literature, it is in vogue to study the outburst properties using a so-called `q'-diagram where hardness is 
plotted with intensity throughout the outburst (Paper-I and references therein). 
%\citep[][and references therein]{Nandi12}.
One of the disadvantages of this is that there is no unique definition of hardness and the energy ranges used in 
the numerator and denominator are rather arbitrary and may be mixtures of soft and Comptonized photons depending 
on the mass of the black holes. In a TCAF model, we directly extract the accretion rate components and visualize 
what happens to the object during the outburst (see also, Mandal \& Chakrabarti, 2010). %\citep[see also,]{Mandal10}
Our direct approach gives more physical insight. Particularly interesting is that the
accretion rate ratio plays a very vital role in defining the spectral state transitions. 

Earlier we showed (DMC14) how spectral state transitions occur due to variation of Keplerian and 
sub-Keplerian rates during 2010-11 outburst of Galactic BHC GX 339-4. We showed variation of physical 
parameters, such as accretion rates, ARR, and shock location and certain behaviors of these parameters 
at the time of state transitions. 
From the variation of Keplerian rate in both these cases, we are now convinced that an 
outburst is triggered due to a sudden rise in viscosity. Viscous time scale brought much of Keplerian disk 
matter in a time scale of about eight days (MJD 55417 to 55425) in H~1743-322. Drainage took longer time as 
this matter is entrained with sub-Keplerian halo. It would be interesting to check if flow dynamics 
of other outbursting sources also follow a similar trend. Prediction of QPO frequency from TCAF 
solution fitted shock parameters ($X_s$ \& $R$), and comparative study with POS model solution (Paper-I) 
will be published elsewhere.

\section*{Acknowledgments}

S. Mondal acknowledges the support of CSIR-NET scholarship.

\begin{figure}[h]
\vspace {0.5cm}
%\epsscale{.98}
%\plotone{fig1.eps}
\centering{
\includegraphics[scale=0.6,angle=0,width=14truecm]{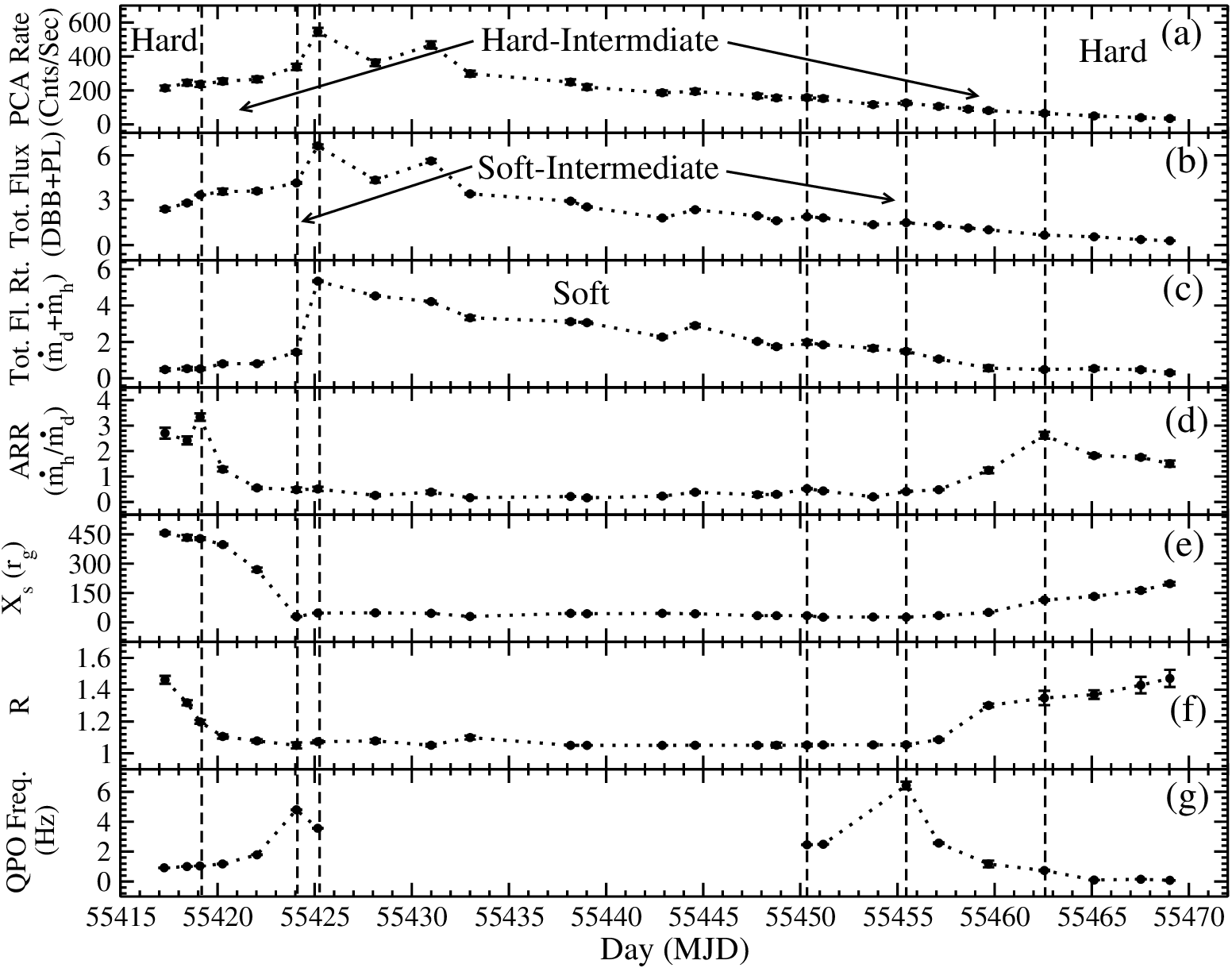}}
\caption{Variation of (a) $2-25$~keV PCA count rates (cnts/sec), (b) combined disk black body (DBB) and 
power-law (PL) model fitted total spectral flux in $2.5-25$~keV range, (c) TCAF model fitted total accretion 
rate (sum of Keplerian disk rate $\dot{m_d}$ and sub-Keplerian halo rate $\dot{m_h}$) in the $2.5-25$~keV energy 
band, (d) ARR (i.e., ratio between halo and disk rates) with day (MJD) for the 2010 outburst of H~1743-322 
are shown. In bottom three panels, variations of observed QPO frequency in Hz (g) and shock location (e), 
compression ratio (f) are shown. Vertical dashed lines are drawn where we believe transitions of different 
spectral states have actually taken place.}
\label{fig1}
\end{figure}

\begin{figure}
\vspace {0.5cm}
%\epsscale{.98}
%\plotone{fig2.eps}
\centering{
\includegraphics[scale=0.6,angle=0,width=14truecm]{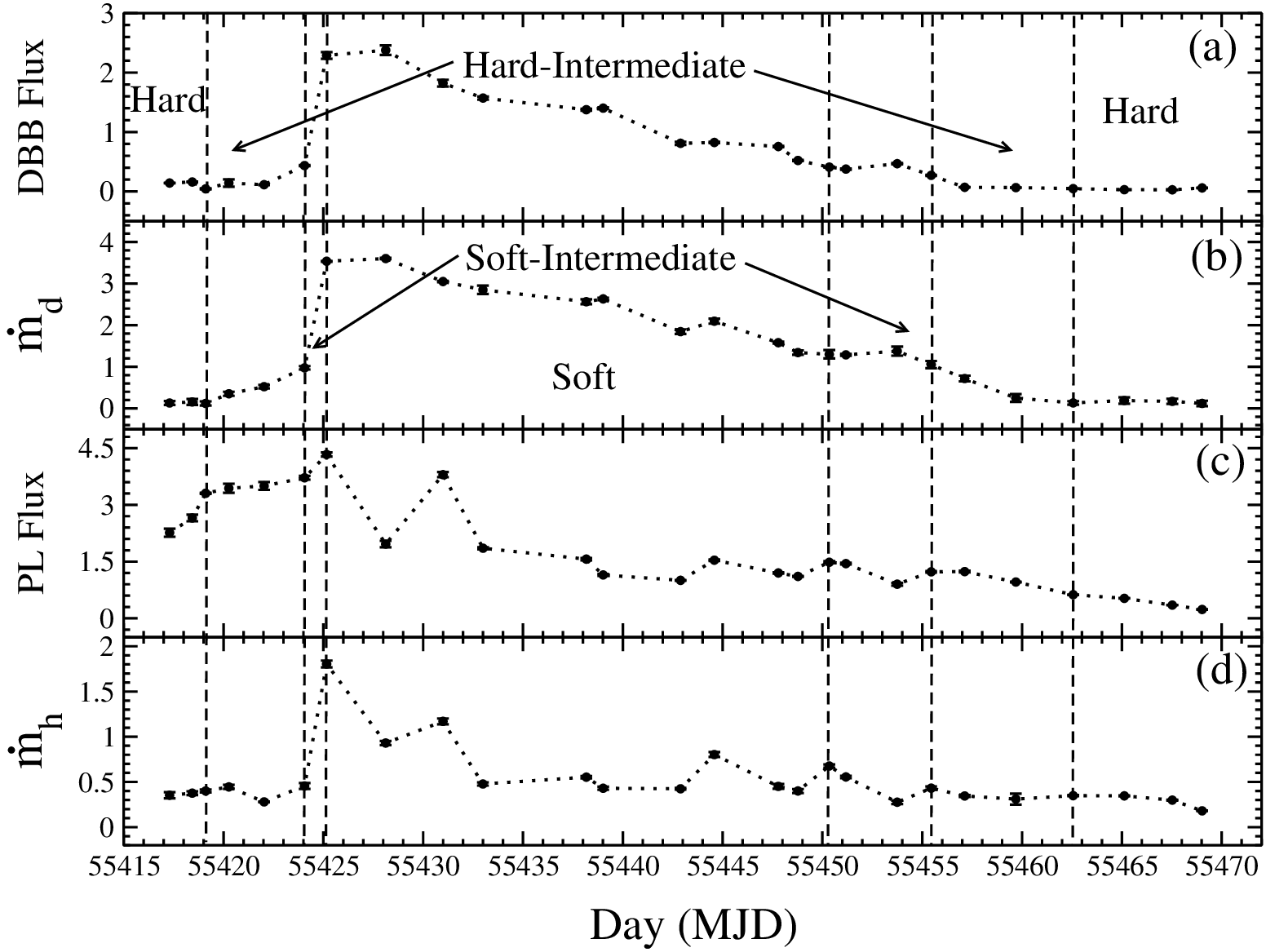}}
\caption{Variation of combined disk black body (DBB) and power-law (PL) model fitted DBB
spectral flux in $2.5-25$~keV energy range (a) and variation of TCAF model fitted Keplerian
disk rate ($\dot{m_d}$). Though models are different, these variations appear to be very similar. 
This is a consistency check for the TCAF solution. In (c), PL spectral flux in same energy range 
and in bottom panel (d), variation of TCAF model fitted sub-Keplerian halo rate ($\dot{m_h}$) 
in the same energy band are shown. Though models are different, variations of DBB flux and $\dot{m_d}$ 
or PL flux and $\dot{m_h}$ appear to be very similar.}
\label{fig2}
\end{figure}

\begin{figure}
\vbox{
\vskip 0.7cm
\centerline{
\includegraphics[scale=0.6,angle=0,width=10truecm]{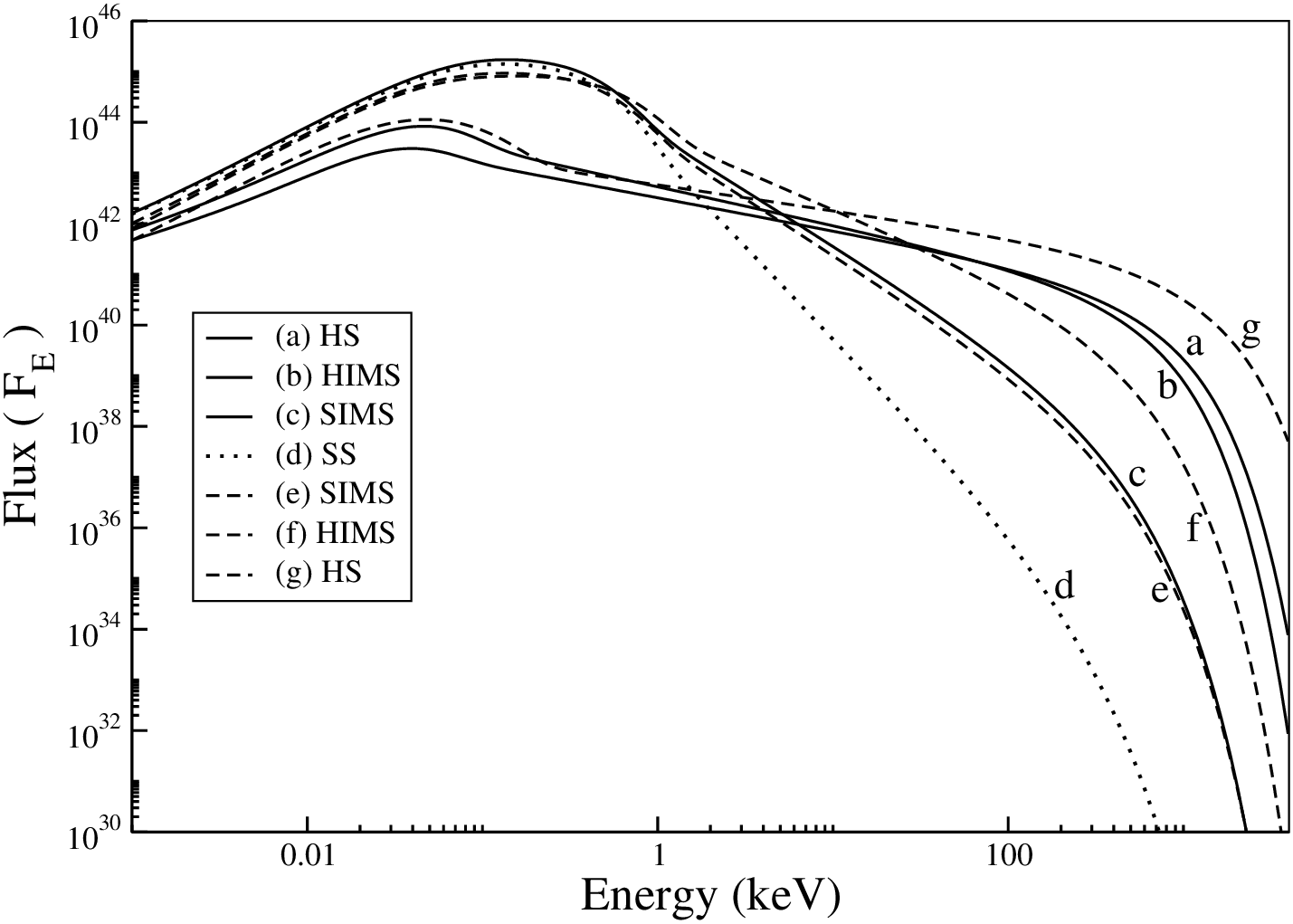}}
\caption{Unabsorbed TCAF model generated spectra for four different spectral states, which were used to fit
the spectra as marked in Table 1. Solid-line plots (a-c) are from the rising phase and dashed-line plots (e-g) 
are from the declining phase of the outburst. Dotted plot (d) shows the spectrum of the soft state. 
Flux ($F_E$) is in units of $photons~cm^{-2}~sec^{-1}~keV^{-1}$. }
\label{fig3}}
\end{figure}

\begin{figure}
\vbox{
\vskip -0.5cm
\centerline{
\includegraphics[scale=0.6,angle=0,width=14truecm]{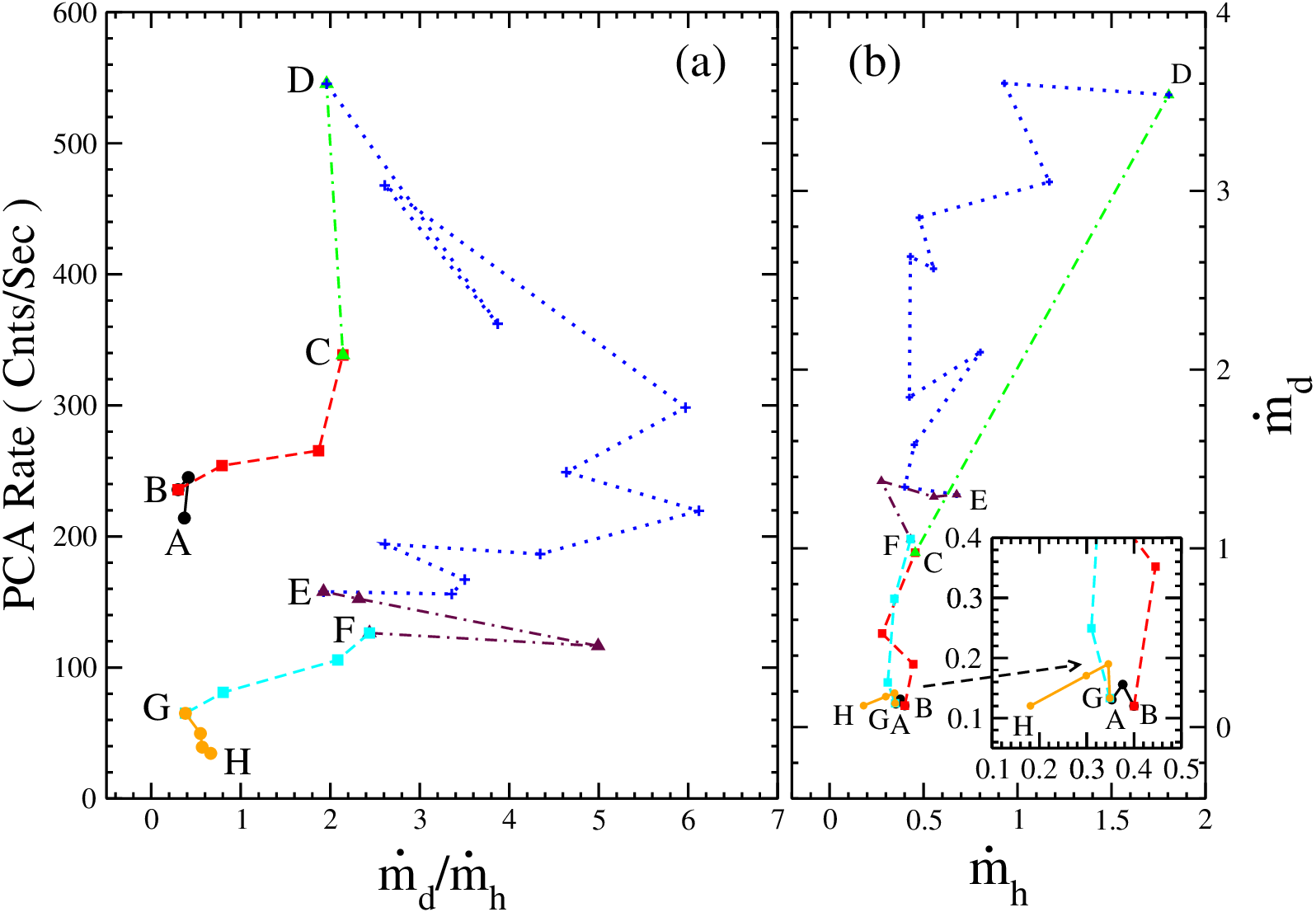}}
\caption{(a) Variation of PCA count rate as a function of 1/ARR and (b) disk rate as a function of  
halo rate (b) for the whole outburst. Transitions from hard to hard intermediate (rising) and vice versa (declining)
takes place when ARR is separetely maximum. Soft state begins with the highest value of individual rates.
Duration of hard intermediate states in both phases occur at a similar value of ARR. 
For ARR $\leq 0.5$, QPOs are sporadic or absent.}
%%For ARR $\lsim 0.5$, QPOs are sporadic or absent.}
\label{fig4}}
\end{figure}

%\begin{figure}%[h]
%%\vskip -0.01cm
%\centerline{
%\includegraphics[scale=0.6,angle=270,width=4truecm]{fig4a.ps}
%\includegraphics[scale=0.6,angle=270,width=4truecm]{fig4g.ps}} 
%\centerline{
%\includegraphics[scale=0.6,angle=270,width=4truecm]{fig4b.ps}
%\includegraphics[scale=0.6,angle=270,width=4truecm]{fig4f.ps}} 
%\centerline{
%\includegraphics[scale=0.6,angle=270,width=4truecm]{fig4c.ps}
%\includegraphics[scale=0.6,angle=270,width=4truecm]{fig4e.ps}}
%\centerline{\includegraphics[scale=0.6,angle=270,width=5truecm]{fig4d.ps}}
%\caption{%{\bf (a-c,e-g)} 
%TCAF model fitted $2.5-25$~keV PCA spectra with variation of $\Delta \chi$, 
%selected from the seven different spectral states whose results are presented in Table 1 are shown.
%Left panel spectra (a-c) are from rising phase and right panel spectra (e-g) are from declining phase 
%of the outburst. Left and Right panels show the hard, hard-intermediate, and soft-intermediate spectra
%for the rising and declining phases respectively. In the bottom panel, (d), the soft state spectrum is shown.}
%\label{fig4}
%\end{figure}

%\begin{figure}%[h]
%\vbox{
%\vskip -0.2cm
%\centerline{
%\noindent{\bf Figure 4d.} Same as Fig. 4(a-c,e-g), only for the soft state spectrum.
%\label{fig4}}
%\end{figure}

%\newpage
%\begin{table}[h]
\begin{table}
%\vskip -2.5cm
\addtolength{\tabcolsep}{-2.50pt}
%\scriptsize
\centering
%\centering{\large \bf Appendix A}
\vskip 0.2cm
\centerline {Table 1}
\centerline {2.5-25 keV TCAF Model Fitted Parameters with QPOs}
\vskip 0.2cm
\begin{tabular}{lcccccccccc}
\hline
\hline
Obs.&Id&UT&MJD&$\dot{m_d}$&$\dot{m_h}$&ARR&$X_s$&R&QPO&$\chi^2/DOF$\\
    &  & & &($\dot{M}$$_{Edd}$)&($\dot{M}$$_{Edd}$)& &($r_g$)& &(Hz)&   \\
 (1)&  (2)  & (3)  & (4)& (5) & (6) & (7) &  (8) & (9) & (10) & (11) \\
\hline

 1    &X-01-00&09/08/10&55417.29&$0.131\pm0.036$&$0.353\pm0.023$&$2.702\pm0.218$&$456.0\pm6.15$&$1.462\pm0.025$&$0.919\pm0.004$&59.28/42 \\
 2$^a$&Y-01-00&10/08/10&55418.43&$0.156\pm0.052$&$0.376\pm0.021$&$2.415\pm0.157$&$433.1\pm13.7$&$1.318\pm0.016$&$1.002\pm0.003$&53.66/42 \\
 3    &Y-02-01&11/08/10&55419.10&$0.120\pm0.051$&$0.400\pm0.016$&$3.333\pm0.146$&$427.9\pm6.56$&$1.198\pm0.012$&$1.045\pm0.008$&62.72/42 \\
 4    &Y-02-00&12/08/10&55420.26&$0.352\pm0.049$&$0.445\pm0.023$&$1.265\pm0.090$&$397.4\pm4.65$&$1.106\pm0.013$&$1.174\pm0.002$&67.43/42 \\
 5$^b$&Y-02-03&14/08/10&55422.03&$0.524\pm0.041$&$0.279\pm0.014$&$0.535\pm0.043$&$269.6\pm10.4$&$1.078\pm0.010$&$1.789\pm0.017$&70.23/42 \\
 6    &Y-03-01&16/08/10&55424.06&$0.976\pm0.042$&$0.456\pm0.033$&$0.467\pm0.090$&$29.12\pm1.56$&$1.050\pm0.019$&$4.796\pm0.022$&42.86/42 \\
 7$^c$&Y-04-00&17/08/10&55425.16&$3.537\pm0.011$&$1.805\pm0.038$&$0.510\pm0.083$&$47.66\pm1.97$&$1.073\pm0.009$&$3.558\pm0.024$&74.73/43 \\
 8    &Y-05-00&20/08/10&55428.12&$3.602\pm0.015$&$0.931\pm0.023$&$0.258\pm0.044$&$48.56\pm2.39$&$1.078\pm0.011$&$     -----   $&75.56/42 \\
 9    &Y-07-00&22/08/10&55430.99&$3.050\pm0.016$&$1.169\pm0.032$&$0.383\pm0.065$&$46.29\pm2.26$&$1.051\pm0.008$&$     -----   $&65.02/44 \\
10    &Y-06-01&24/08/10&55432.99&$2.850\pm0.099$&$0.477\pm0.017$&$0.168\pm0.012$&$29.98\pm1.91$&$1.099\pm0.012$&$     -----   $&50.67/42 \\
11    &Y-10-00&30/08/10&55438.17&$2.565\pm0.061$&$0.553\pm0.013$&$0.216\pm0.010$&$46.02\pm2.55$&$1.050\pm0.002$&$     -----   $&59.59/45 \\
12$^d$&Y-11-00&31/08/10&55439.01&$2.633\pm0.028$&$0.430\pm0.017$&$0.163\pm0.030$&$44.34\pm3.23$&$1.050\pm0.002$&$     -----   $&69.60/45 \\
13    &Y-13-00&03/09/10&55442.89&$1.846\pm0.055$&$0.425\pm0.019$&$0.230\pm0.026$&$46.25\pm0.17$&$1.050\pm0.002$&$     -----   $&79.29/45 \\
14    &Y-15-00&05/09/10&55444.58&$2.098\pm0.060$&$0.803\pm0.027$&$0.383\pm0.028$&$44.09\pm0.07$&$1.051\pm0.003$&$     -----   $&59.33/45 \\
15    &Y-18-00&08/09/10&55447.78&$1.579\pm0.022$&$0.451\pm0.025$&$0.286\pm0.079$&$34.27\pm0.06$&$1.051\pm0.005$&$     -----   $&34.08/44 \\
16    &Y-19-00&09/09/10&55448.77&$1.343\pm0.052$&$0.399\pm0.021$&$0.298\pm0.037$&$34.21\pm0.65$&$1.051\pm0.015$&$     -----   $&51.43/45 \\
17    &Y-20-00&11/09/10&55450.34&$1.301\pm0.101$&$0.676\pm0.018$&$0.519\pm0.019$&$34.21\pm1.48$&$1.052\pm0.008$&$2.454\pm0.022$&33.70/44 \\
18$^e$&Y-20-01&12/09/10&55451.17&$1.288\pm0.024$&$0.555\pm0.013$&$0.431\pm0.023$&$26.52\pm1.16$&$1.053\pm0.004$&$2.489\pm0.018$&43.51/45 \\
19    &Y-21-01&14/09/10&55453.74&$1.377\pm0.108$&$0.276\pm0.019$&$0.201\pm0.026$&$27.54\pm1.51$&$1.053\pm0.005$&$     -----   $&39.90/45 \\
20    &Y-22-01&16/09/10&55455.44&$1.052\pm0.088$&$0.432\pm0.018$&$0.410\pm0.026$&$26.40\pm1.78$&$1.054\pm0.008$&$6.417\pm0.252$&41.57/42 \\
21$^f$&Y-23-01&18/09/10&55457.12&$0.718\pm0.069$&$0.345\pm0.010$&$0.480\pm0.018$&$34.65\pm2.24$&$1.086\pm0.006$&$2.569\pm0.044$&42.18/45 \\
22    &Y-24-01&20/09/10&55459.68&$0.249\pm0.043$&$0.310\pm0.020$&$1.243\pm0.110$&$50.81\pm2.79$&$1.301\pm0.011$&$1.172\pm0.226$&83.71/42 \\
23$^g$&Y-25-01&23/09/10&55462.56&$0.133\pm0.032$&$0.349\pm0.015$&$2.620\pm0.133$&$114.4\pm3.21$&$1.348\pm0.045$&$0.741\pm0.046$&52.27/42 \\
24    &Y-28-00&26/09/10&55465.12&$0.190\pm0.034$&$0.346\pm0.011$&$1.820\pm0.068$&$132.3\pm5.63$&$1.369\pm0.027$&$0.102\pm0.003$&41.33/42 \\
25    &Y-26-02&28/09/10&55467.52&$0.171\pm0.062$&$0.299\pm0.010$&$1.755\pm0.064$&$163.0\pm7.82$&$1.429\pm0.052$&$0.149\pm0.005$&36.72/42 \\
26    &Y-28-01&30/09/10&55469.01&$0.120\pm0.065$&$0.181\pm0.013$&$1.504\pm0.122$&$197.9\pm8.46$&$1.471\pm0.054$&$0.079\pm0.002$&38.72/42 \\

\hline
\end{tabular}
\noindent{
\leftline {Here X=95368-01, Y=95360-14 indicate the initial part of the observation Ids, and UT date in dd/mm/yy format.}
\leftline {$\dot{m_h}$, and $\dot{m_d}$ represent TCAF model fitted sub-Keplerian (halo) and Keplerian (disk) rates in Eddington rate respectively.}
\leftline {$X_s$ (in Schwarzchild radius $r_g$), and $R$ are the TCAF model fitted shock location and compression ratio values respectively. }
%\leftline {$h_{shk}$ (in $r_g$) and $T_{shk}$ (in K) are the shock height and temperature values derived from Eqs. 4 \& 5  respectively.}
\leftline {Here, frequency of the pricipal QPO in Hz are only presented. DOF means degrees of freedom of the model fit.}
%\leftline {$^*$ observed spectra are fitted with out addition of Iron Line of $\sim 6.5$ keV.}
}

\end{table}

%%%%%%%%%%%%%%%
\end{document}